\documentclass[a4paper,11pt]{article}
\usepackage{jinstpub} 
\usepackage{lineno}
\usepackage{setspace}

\title{\boldmath Investigation of low gain avalanche detectors exposed to proton fluences beyond 10$^{15}$ n$_\mathrm{eq}$cm$^{-2}$}

\author[a,1]{J. Sorenson,}
\author[a]{M. R. Hoeferkamp,}
\author[b]{G. Kramberger,}
\author[a]{S. Seidel,\note{Corresponding author.}}
\author[a]{J. Si}
\affiliation[a]{Department of Physics and Astronomy, University of New Mexico, Albuquerque, NM 87131, USA}
\affiliation[b]{Department of Experimental Particle Physics, Jozef Stefan Institute, Jamova 39, 1000 Ljubljana, Slovenia}

\emailAdd{sorensonj@unm.edu}

\abstract{Low gain avalanche detectors (LGADs) deliver excellent timing resolution, which can mitigate mis-assignment of vertices associated with pileup at the High Luminosity LHC and other future hadron colliders. The most highly irradiated LGADs will be subject to $2.5 \times10^{15} \mathrm{n}_\mathrm{eq} \mathrm{cm}^{-2}$ of hadronic fluence during HL-LHC operation; their performance must tolerate this. Hamamatsu Photonics K.K. and Fondazione Bruno Kessler LGADs have been irradiated with 400 and 500 MeV protons respectively in several steps up to $1.5 \times10^{15} \mathrm{n}_\mathrm{eq} \mathrm{cm}^{-2}$. Measurements of the acceptor removal constants of the gain layers, evolution of the timing resolution and charge collection with damage, and inter-channel isolation characteristics, for a variety of design options, are presented here.}


\arxivnumber{} 

\begin{document}
\maketitle
\flushbottom

\section{Introduction}
\label{sec:intro}

The High Luminosity LHC (HL-LHC) upgrade is expected to come online in 2029 and will increase the number of interactions by an order of magnitude, up to about 200 interactions per bunch crossing. This increases the pileup and complicates the track reconstruction.  The temporal distribution of proton-proton collisions within a bunch crossing will span less than 180 ps. In order to reduce the likelihood of track assignment to the wrong collision vertex, experiments at the HL-LHC must have timing resolution significantly smaller than 180 ps. ATLAS is installing the High Granularity Timing Detector (HGTD) \cite{HGTD} in the forward region, which should provide a timing resolution of about 30 ps at the start of the HL-LHC run and about 50 ps after receiving a fluence of $2.5 \times10^{15} \mathrm{n}_\mathrm{eq} \mathrm{cm}^{-2}$. CMS is installing the Endcap Timing Layer (ETL) with similar timing goals \cite{ETL}.  Both the ETL and HGTD will be made of low gain avalanche detectors (LGADs)~\cite{LGAD1,LGAD2}. LGADs have excellent timing resolution. The LGADs' charge collection should be above 10 fC at the start of the HL-LHC run and be above 4 fC after receiving a fluence of $2.5 \times10^{15} \mathrm{n}_\mathrm{eq} \mathrm{cm}^{-2}$. Improvement of their radiation hardness is the goal of this study, which compares prototypes by two manufacturers and for a variety of design options.

An irradiation campaign was carried out on the second run of LGAD prototypes produced by Hamamatsu Photonics K.K. (HPK2) and the fourth run of LGAD prototypes produced by Fondazione Bruno Kessler (FBK4) with a variety of design options. Measurements of the devices' timing resolution, gain layer acceptor removal constant, and inter-electrode isolation are reported here. Fluence was applied to the FBK4 devices at the Los Alamos Neutron Science Center (LANSCE) with 500 MeV protons, and to the HPK2 devices at the Fermilab Irradiation Test Area (ITA) with 400 MeV protons. The radiation hardness factors for 400 MeV and 500 MeV protons are 0.83 and 0.78 respectively so there should be negligible differences in the damage between the two irradiations. The irradiations took place at room temperature. Within a few hours after irradiation, all devices were transferred to freezers where they were maintained at -25$^\circ$C continuously to prevent unintentional annealing. Prior to the start of the measurement process, every device was subjected to a standard annealing regimen of 60$^\circ$C for 80 minutes. 

LGADs are thin sensors that have an $n^{++}-p^+-p-p^{++}$ structure. The sensors are operated under reverse bias with their $p$ bulk fully depleted. The $p^+$ layer is referred to as the gain layer or charge multiplication layer. When fully depleted, the electric field across the gain layer can exceed $4\times 10^5$ $ \mathrm{V/cm}$ \cite{breakdown}. Charged particles from LHC collisions passing through  the sensor will produce electron-hole pairs. The electrons undergo charge multiplication via impact ionization in the gain layer. Their intrinsic gain permits these devices to operate with active volumes only several tens of microns thick while achieving charge collection levels greater than that of thicker conventional sensors.  The small thickness allows a short collection time with a fast rising edge that results in their precise timing characteristic.

\section{Features of the Prototypes and Determination of the Applied Fluence}

Prototypes of three structures were produced by HPK2 using epitaxial silicon grown on a Czochralski substrate; these are single LGADs, 2x2 ("quad") LGAD arrays, and associated p-i-n diodes (“pin”). All have 50 $\mathrm{\mu m}$ active layer thickness, 200 $\mathrm{\mu m}$ total thickness, and a single guard ring. The $n^{++}$ electrode has dimensions 1.3 mm x 1.3 mm. The $p^+$ gain layer profile is gaussian with approximately 0.7 $\mu$m thickness and 1.8 $\mathrm{\mu m}$ depth. The quad arrays have the same features and an inter-electrode separation of 50~$\mathrm{\mu m}$ as well as two design options for the distance from the active area to the edge (300 $\mathrm{\mu m}$ and 500 $\mathrm{\mu m}$). The singles and quads were produced with four different options on gain layer doping profile. Profile 1 corresponds to the highest dose, and Profile~4 corresponds to the lowest dose. The difference in doping levels between Profiles 1 and 4 is roughly 6\%.  Dopant concentrations of only a few percent difference have previously been shown to lead to large differences in gain \cite{gainDiff}. The pin diodes have the same geometry as the LGADs but lack the gain layer. Pin structures are useful for assessing the breakdown characteristic, as breakdown in a pin is indicative of breakdown in the bulk, typically at the guard ring where the field lines are focused. The pre-irradiation gain layer depletion voltage, $V_\mathrm{gl,0}$, full depletion voltage  $V_\mathrm{fd,0}$, and some specifications of the HPK2 prototypes are shown in Table \ref{tab:hpk}.

\begin{table}[htbp]
\centering
\caption{Specifications of the HPK2 LGAD prototypes used in this study. All prototypes have an active thickness of 50 $\mathrm{\mu m}$ and a 1.8 $\mu$m deep gain layer. The initial gain layer and full depletion voltages were measured at the University of New Mexico and are representative of all devices from the wafer.\label{tab:hpk}}
\smallskip
\begin{tabular}{c|c|c|c} 
\hline
HPK2 Wafer ID&Gain Layer Doping Profile & $V_\mathrm{gl,0} [\mathrm{V}]$ & $V_\mathrm{fd,0} [\mathrm{V}]$ \\
\hline
25 & 1 & 53.0 & 55.0\\
31 & 2 & 52.0 & 54.0\\
36, 37 & 3 & 49.5 & 52.0\\
42, 43 & 4 & 49.0 & 51.0 \\
\hline
\end{tabular}
\end{table}

The FBK4 LGADs also explore a range of parameters. They have a slightly larger active thickness of 55 $\mathrm{\mu m}$ and 500 $\mathrm{\mu m}$ total thickness, 1.3 mm x 1.3 mm electrode size, multiple guard rings, and carbon co-implantation in the gain layer. The FBK4 quads explore options in the electrode isolation. Both of the quad options are produced with a p-stop and junction terminating extension (JTE). The quad labeled T10 has a guard ring between the pixels whereas the quad labeled T9 does not have a ground between the pixels. This ideally makes the T10 quad safer for breakdowns. The size of the inter-electrode gap in the T9 and T10 quads is slightly different and approximately 50~$\mathrm{\mu m}$. The carbon co-implantation has shown improved radiation hardness to neutrons and protons \cite{carbonEffect, UFSD_c_table}. These prototypes are produced with variation in the boron concentration and the carbon concentration in the gain implant. The depth of the gain implant is controlled by the energy of the implanter and is produced with two options, shallow and deep. The depth of the shallow option is less that 1 $\mu$m, and that for the deep option is approximately 2 $\mu$m. The width of the gain layer is controlled by the thermal load in the diffusion process. A high thermal load corresponds to a wider gain layer while a low thermal load corresponds to a relatively thinner gain layer. With high carbon diffusion and high boron diffusion (CH-BH), the sensor would have both carbon and boron implanted and then both implants are subjected to a high thermal load resulting in both implants being relatively wider. Likewise, with low carbon diffusion and low boron diffusion (CL-BL), the sensor would have both carbon and boron implanted and then both implants are subjected to a low thermal load, resulting in both implants being relatively thin. With high carbon diffusion and low boron diffusion (CH-BL), the sensor would have carbon implanted, then be subjected to a high thermal load, then have boron implanted, and then be subjected to a low thermal load, resulting in a wider carbon implant and a thin boron implant. The diffusion processes are described in more detail in \cite{diffusion}. The pre-irradiation gain layer depletion voltage, full depletion voltage, and some specifications of the FBK4 prototypes are shown in Table \ref{tab:fbk}.

\begin{table}[htbp]
\centering
\caption{Specifications of the FBK4 LGAD prototypes used in this study. The thickness is 55 $\mathrm{\mu m}$ for all prototypes. Wafers 1-9 have shallow gain layers and wafers 12-18 have deep gain layers. Relative concentrations of boron and carbon are provided. The diffusion column indicates the thermal load applied to the carbon and boron implants. A larger thermal load corresponds to a wider implant. The initial gain layer and full depletion voltages were measured at the University of New Mexico and are representative of all devices from the wafer. \label{tab:fbk}}
\smallskip
\begin{tabular}{c|c|c|c|c|c|c} 
\hline
\multicolumn{1}{|p{1.5cm}|}{\centering FBK4 Wafer ID}&\multicolumn{1}{|p{2cm}|}{\centering Gain Layer Depth}&\multicolumn{1}{|p{2.5cm}|}{\centering Relative Boron Concentration}&\multicolumn{1}{|p{2.5cm}|}{\centering Relative Carbon Concentration}&Diffusion & $V_\mathrm{gl,0} [\mathrm{V}]$ & $V_\mathrm{fd,0} [\mathrm{V}]$\\
\hline
1  & Shallow & 0.98 & 0.6 & CH-BL & 21.5 & 23.0 \\
2  & Shallow & 1.02 & 1 & CH-BL & 22.0 & 23.5 \\
5  & Shallow & 1.04 & 1 & CH-BL & 22.5 & 24.0 \\
9  & Shallow & 1.06 & 1 & CH-BL & 22.5 & 24.5 \\
12 & Deep & 0.77 & 0.6 & CH-BH & 50.5 & 51.5 \\
16 & Deep & 0.81 & 0.6 & CL-BL & 48.0 & 49.0 \\
18 & Deep & 0.93 & 0.6 & CL-BL & 48.5 & 49.5 \\
\hline
\end{tabular}
\end{table}

Every prototype inserted into the beamline was mounted on a radiation-hard fiberglass laminate, G-10, board and accompanied on the axis of the same board by a thin high purity aluminum foil of area 1 cm$^2$. The beam to which the sensors were exposed had a profile that is approximately gaussian.  Dosimetry via gamma spectrometry of the aluminum foil provides a measure of the average fluence received over the foil's area.  However the fluence received by any particular sensor depended upon its precise position within the beam.  The uncertainty on the fluence that each received is a combination of the uncertainty on the aluminum foil dosimetry and the uncertainty on the location of the sample within the beam spot.

The activation of the aluminum foils was measured with the gamma spectrometer, and then the Monte Carlo N-Particle (MCNP) simulation \cite{MCNP} was used to infer the fluence of 400 MeV or 500 MeV protons that had induced it.  In the case of the 400 MeV samples, the MCNP simulation was validated to 10\% uncertainty by a separate calculation using proton-on-aluminum cross section values reported in Reference \cite{Morgan}. The uncertainty on the aluminum dosimetry due to counting statistics is less than 10\% for all samples. The 400 MeV and 500 MeV proton fluences are scaled to the 1-MeV neutron equivalent fluence ($\mathrm{n}_\mathrm{eq} \mathrm{cm}^{-2}$) using the radiation hardness factors listed in the introduction. 

The beam profile was characterized with arrays of OSRAM BPW34F p-i-n diodes positioned at the upstream and downstream ends of the beamline.  These diodes function effectively up to saturation around $10^{15}  \mathrm{n}_\mathrm{eq} \mathrm{cm}^{-2}$.  The diode voltage is calibrated to fluence in Reference \cite{Hoeferkamp}. The average dose measured over the central 1 cm$^2$ region of the diode array was shown to agree well with the dose obtained using gamma spectroscopy on the Al foil. 

The diode arrays demonstrated that the beam profile was approximately gaussian, displaced ~5 mm from the axis of the G-10 boards on which the samples were centered, and of FWHM approximately 1 cm. While all sensors overlapped aluminum foils, the small radius of the beam means that sensors near the beam axis received different integrated fluence than sensors farther from the beam axis.  

The uncertainty on each sensor's integrated fluence, due to the beam's gaussian profile and the offset, is 66\%. The total uncertainty on the measured fluence is the quadrature sum of the three contributions above, which is 68\%.

\section{Measurements and Results}

\subsection{Leakage Current and Capacitance}

\begin{figure}[htbp]
	\centering
	\includegraphics[width=1\textwidth]{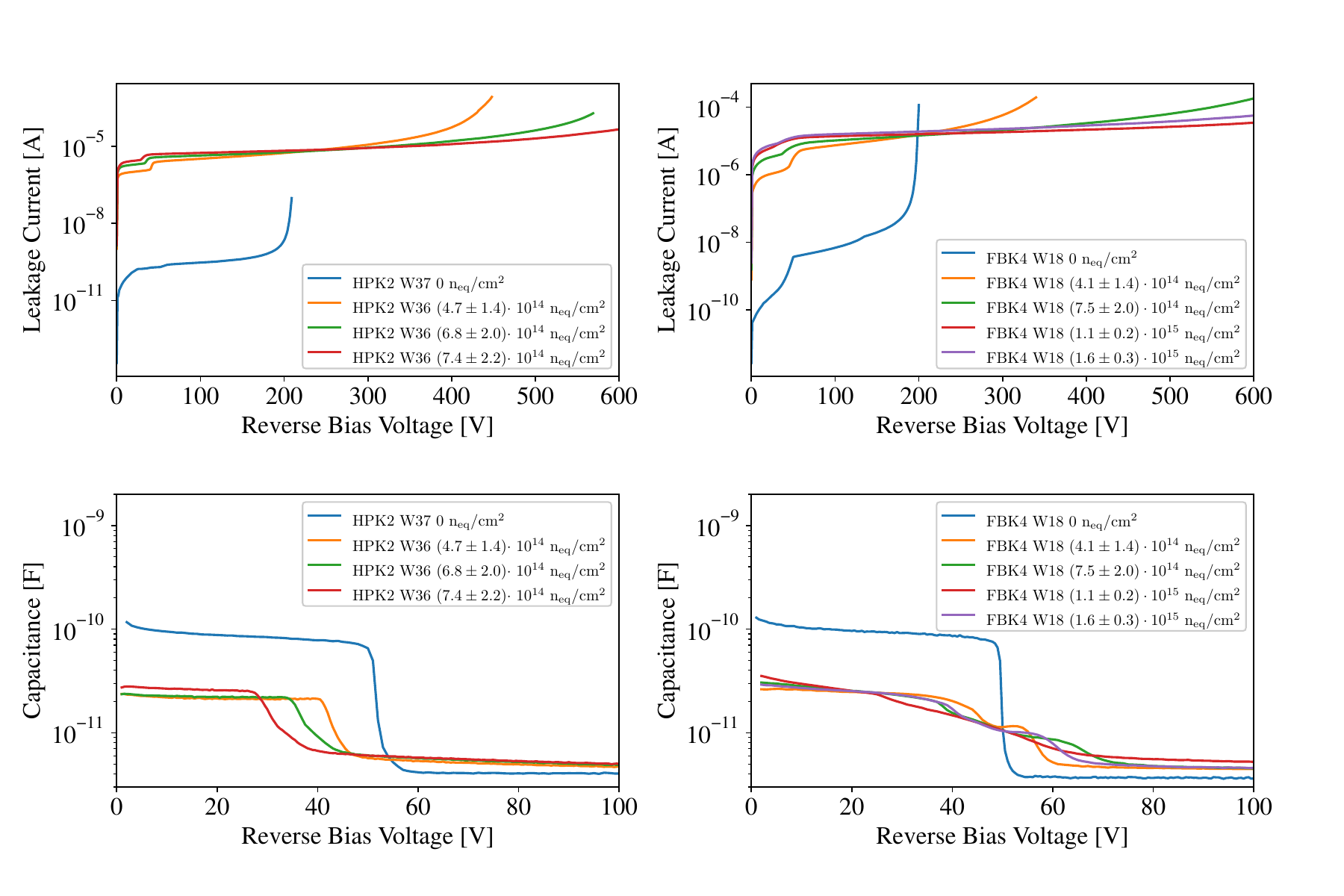}
	\caption{Leakage current versus bias voltage of HPK2 W36 (upper left) and FBK4 W18 (upper right), and capacitance versus bias voltage of HPK2 W36 (lower left) and FBK4 W18 (lower right). These plots are representative of all measured sensors. The applied proton fluences are shown in the embedded text.  \label{fig:IVandCV}}
\end{figure}

Measurements of leakage current versus bias voltage (IV) and capacitance versus bias voltage (CV) were carried out before and after exposure to radiation. The IV was measured in 1 V steps until the current exceeded $2\times10^{-4}$  A or the bias voltage exceeded 600 V. Biasing above 600 V can lead to single event burnout which renders the LGAD inoperable \cite{singleEventBurnout}. Figure \ref{fig:IVandCV} shows sample IV curves for a set of devices from FBK4 W18 and HPK2 W36 which are representative of all sensors. The CV was measured in 0.5 V increments up to 100 V at 10 kHz. Both the CV and IV measurements were performed on a thermal chuck with a temperature of $20.0 \pm 0.5 ^\circ \mathrm{C} $. The uncertainty in temperature contributes 3\% to the error in the leakage current measurement.

\begin{figure}[htbp] 
	\centering
	\includegraphics[width=.45\textwidth]{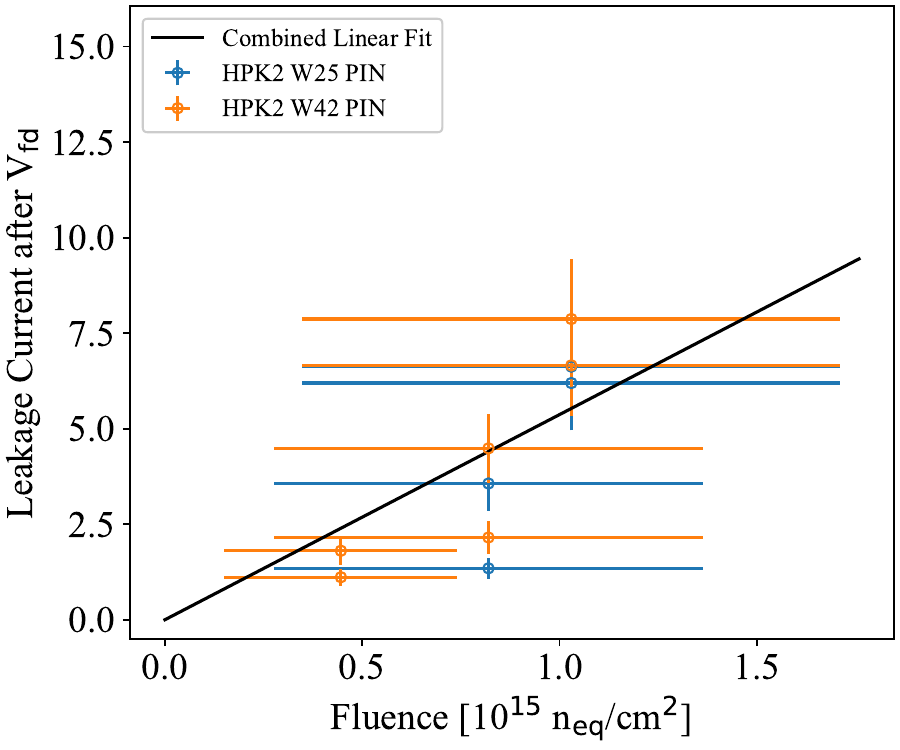}
	\qquad
	\includegraphics[width=.45\textwidth]{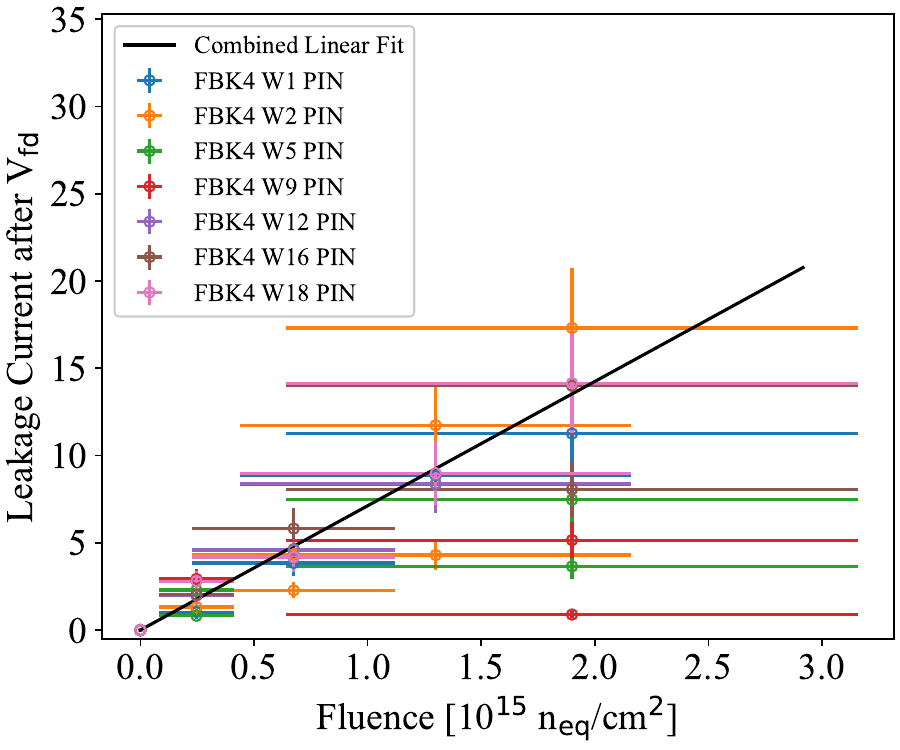}
	\caption{The leakage current 5 V above full depletion versus proton fluence for HPK2 (left) and FBK4 (right) pins. The expression $\Delta I = \alpha V \phi$ is fit to the data in both cases to extract the damage constant. The results of these fits are shown in Table \ref{tab:damageConstant}. \label{fig:damageConstant}}
\end{figure}

A  feature of silicon sensors is a linear increase in leakage current $\Delta I$ with bulk damaging fluence. The proportionality between the fluence $\phi$ and leakage current is given by 

\begin{equation}
	\Delta I = \alpha \mathbb{V} \phi
\end{equation}
where $\mathbb{V}$ is the volume of the sensor and $\alpha$ is the damage constant \cite{damageConstant}. The unirradiated leakage current, $I_0$, is several orders of magnitude lower than the irradiated leakage current, $I$, so the replacement $\Delta I = I- I_0\approx I$ is made. For LGADs, this expression requires an additional factor to account for the gain, \textit{g}, and can be written as $I = g \alpha \mathbb{V} \phi$. The leakage current of the pin sensors is extracted 5 V above full depletion. The observed leakage current measured at this bias voltage is used to determine $\alpha$, for fluences obtained by the method described in Section 2. This fit is shown in Figure \ref{fig:damageConstant} and the fit results are shown in Table \ref{tab:damageConstant}. Since the voltage of full depletion, $V_\mathrm{fd}$, is a function of the sensor's dose, a systematic error is associated with measuring the leakage current above $V_\mathrm{fd}$. The change in leakage current over the range of measured $V_\mathrm{fd}$ in a pin sensor is used to estimate this error conservatively at 20\%. The error reported is the total error from the linear fit, which takes into account the systematic error in determining the leakage current and the error in fluence from each measurement, as well as the statistical error on the fit. 


This process was repeated for the LGADs with leakage current extracted 5 V beyond full depletion. At this bias, the sensor is fully depleted but produces little gain so it is nearly comparable to the gainless pin sensors. Prior studies have shown that the damage constant has no dependence on carbon doping \cite{effectOfCarbon}, so there should be little impact from the doping profiles of the sensors in the combined fit. The result for the damage constant from separate combined fits of all the FBK4 LGAD and HPK2 LGAD sensors' leakage currents are also reported in Table \ref{tab:damageConstant}. 

A second step is taken to fine tune the dosimetry of the individual LGAD sensors. Using the calculated average values of $\alpha$ for each manufacturer's set of devices, and each device's particular leakage current $I$ and volume $\mathbb{V}$, a corrected fluence for each LGAD sensor is determined with Equation 3.1. The $\alpha$ values calculated from the pin sensors are used in this calculation because there is no possibility of systematic increase in the value of $\alpha$ from gain. The leakage current of the LGAD is extracted just above full depletion and has a 20\% error from the choice of voltage. The error in the calculated fluence is the combination of the errors in $\alpha$ and the uncertainty in determining the leakage current of the LGAD sensor. These two errors are uncorrelated since the damage constant fit was performed on the pin sensors, and the fluence correction is only applied to the LGADs. This corrected fluence is used for the quad and single LGADs in each section.


\begin{table}[htbp]
	\centering
	\caption{Damage constant \cite{damageConstant} extracted from combined fits of the FBK4 and HPK2 prototypes.\label{tab:damageConstant}}
	\smallskip
	\begin{tabular}{c|cc} 
	\hline
	&HPK2&FBK4\\
	\hline
	pin & $(5.3 \pm 0.7) \cdot 10^{-17} \mathrm{A/cm}$ & $(6.9\pm0.4)\cdot10^{-17} \mathrm{A/cm}$\\
	LGAD& $(5.2 \pm 0.4)\cdot10^{-17} \mathrm{A/cm}$ & $(8.5 \pm 0.6)\cdot10^{-17} \mathrm{A/cm}$\\
	\hline
	\end{tabular}
\end{table}

\subsection{Acceptor Removal Constant}

The voltage required to deplete the gain implant, $V_\mathrm{gi}$, is related to the dopant concentration in the gain layer, N$_A$, by

\begin{equation}
	V_\mathrm{gi} = \frac{q\mathrm{N}_Aw^2}{2\epsilon}.
\end{equation}

Here $q$ is the electron charge, $w$ is the width of the gain implant, and $\epsilon$ is the permittivity of silicon.  The size of the gap between the gain implant and the $n^{++}$ electrode is the gain layer depth, $d$. The voltage required to deplete the gain implant and the gap between the gain implant and the $n^{++}$ electrode is called the gain layer depletion voltage, $V_\mathrm{gl}$. $V_\mathrm{gl}$ is related to the gain layer depth and the gain implant depletion voltage by \cite{vglEquation}

\begin{equation}
	V_\mathrm{gl} \approx V_\mathrm{gi} + V_\mathrm{gap} \approx V_\mathrm{gi}\left(1+2\frac{d}{w}\right).
\end{equation}

The electric field in the gain layer, and therefore the gain, are related to the gain layer depletion voltage. Hadronic fluence will reduce the active dopant concentration via transformation of the boron acceptors into defect complexes no longer acting as acceptors \cite{damageMechanism}. This phenomenon is called acceptor removal and is the primary mechanism by which the gain and consequently the timing performance of LGADs are reduced \cite{radEffectLGAD}.   

Inflections in measurements of the device's bulk capacitance versus applied bias voltage (CV) are used to infer the $V_\mathrm{gl}$ and $V_\mathrm{fd}$ since the capacitance is inversely proportional to the depletion width. Both IV and CV measurements have features that indicate the applied voltage at which the sensor bulk and gain layer reach full depletion. In the IV, there is an easily identifiable inflection. In the CV, the gain layer depletion is indicated when the capacitance begins rapidly to decrease, and the full sensor is depleted when the capacitance subsequently levels off at a higher voltage. This analysis uses an algorithm to identify first the inflection in the IV curve and then searches the range of nearby voltages in the CV curve for the inflection points corresponding to the gain layer depletion and full depletion. We do not use the CV characteristic alone because of the capacitance characteristic's two distinct drops (a double peak behavior) as seen in Figure \ref{fig:IVandCV} for the FBK4 W18 sensor. 

\begin{figure}[htbp]
	\centering
	\includegraphics[width=.6\textwidth]{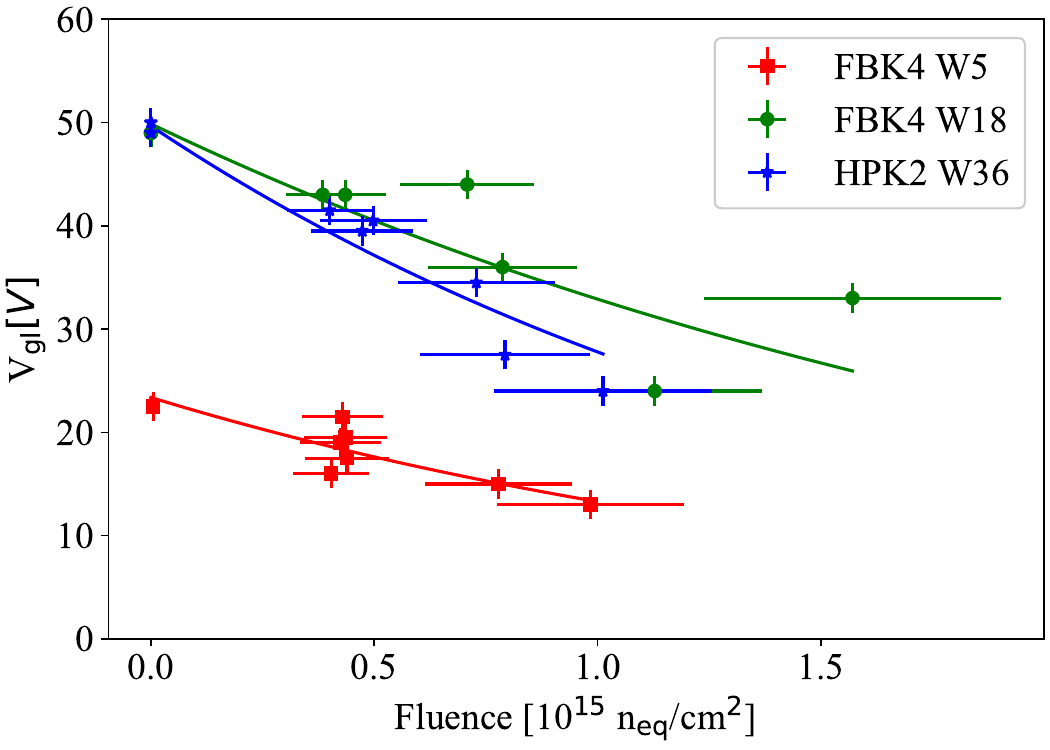}
	\caption{ The gain layer depletion voltage versus applied fluence for FBK4 W5 (carbon co-implant, shallow gain layer), FBK4 W18 (carbon co-implant, deep gain layer) and HPK2 W36 (no carbon implant, deep gain layer). A decaying exponential $V_\mathrm{gl} (\phi)=V_{\mathrm{gl},0}\cdot e^{-c\phi}$ is fit to the data, where $c$ is the acceptor removal constant. The results of these fits are shown in Table \ref{tab:acceptorRemovalConstant}.  \label{fig:acceptorRemoval}}
\end{figure}

Figure \ref{fig:acceptorRemoval} shows the development of $V_\mathrm{gl}$ with fluence for three representative wafers, FBK4 W5, FBK4 W18, and HPK2 W36.  The function $V_\mathrm{gl} (\phi)=V_\mathrm{gl,0} \cdot e^{-c\phi}$ is fit to the gain layer depletion voltage versus fluence to extract the acceptor removal constant, $c$. The vertical error bars represent the uncertainty in determining the voltage where the CV graph has an inflection point, which is related to the sampling frequency. The horizontal error is uncertainty in the dose. Both errors are propagated into the fitting algorithm and contribute to the error on the acceptor removal constant. Table \ref{tab:acceptorRemovalConstant} summarizes extracted values of $c$ for all the sensors studied. The reported values are between 2 and 3 times higher for FBK4 and around 1.5 times higher for HPK2 than those associated with comparable neutron exposures \cite{neutronLGAD}. 

\begin{table}[htbp]
\centering
\caption{Extracted values of the acceptor removal constant for all sensor types studied. The table includes parameters that are relevant to the radiation hardness. \label{tab:acceptorRemovalConstant}}
\smallskip
\begin{tabular}{c|c|c|c|c} 
\hline
Manufacturer &Wafer &\multicolumn{1}{|p{3cm}|}{\centering Carbon Co-Implantation}&\multicolumn{1}{|p{3cm}|}{\centering Gain Layer\\ Depth}&$c$ [$10^{-16}\mathrm{n}_\mathrm{eq}^{-1}\mathrm{cm}^2$] \\
\hline
FBK4  & W1   & Yes & Shallow &  $5.4 \pm 0.9$ \\
FBK4  & W2   & Yes & Shallow &  $5.0 \pm 0.9$ \\
FBK4  & W5   & Yes & Shallow &  $5.7 \pm 1.4$ \\
FBK4  & W9   & Yes & Shallow &  $3.1 \pm 0.7$ \\
FBK4  & W12 & Yes & Deep     &  $6.2 \pm 1.1$ \\
FBK4  & W16 & Yes & Deep     &  $4.5 \pm 0.6$ \\
FBK4  & W18 & Yes & Deep     &  $4.2 \pm 0.8$ \\
HPK2  & W25 & No  & Deep    &  $5.3 \pm 0.3$ \\
HPK2  & W31 & No  & Deep    &  $6.5 \pm 0.9$ \\
HPK2  & W36 & No  & Deep    &  $5.8 \pm 0.5$ \\
HPK2  & W42 & No  & Deep    &  $5.6 \pm 0.3$ \\
\hline
\end{tabular}
\end{table}

\subsection{Timing Resolution and Charge Collection}

The timing resolution and charge collection are measured with a collimated $^{90}$Sr beta source. The device under test (DUT) is measured in triple coincidence with a reference LGAD (REF) above and a SiPM coupled to a scintillator below. Both LGADs are bonded to dedicated preamplifier boards with a gain uncertainty of 10\% \cite{UCSCboard}. The output of those is further amplified by a Particulars AM-02B (35 dB) amplifier before being read out on a Tektronix DPO7254 2.5 GHz 40 GS/s oscilloscope. The measurement is done with the boards in an environmental chamber at -$30 \pm 1.5 ^\circ$C. The time-of-arrival values from the DUT and REF are extracted with a 25\% constant fraction discriminator. The difference between the time-of-arrival values for the two LGADs is a constant that depends on the setup, with width determined by the timing resolution of the LGADs. This width is extracted with a gaussian fit to the difference between the REF and DUT time-of-arrival. The measured width ($\sigma_\mathrm{MEAS}$) is related to the timing resolutions of the REF ($\sigma_\mathrm{REF}$) and DUT ($\sigma_\mathrm{DUT}$) by
\begin{equation}
\sigma_\mathrm{MEAS}^2 = \sigma_\mathrm{REF}^2 +\sigma_\mathrm{DUT}^2.
\end{equation}
The reference is calibrated by a measurement of two identical sensors. In that case, $\sigma_\mathrm{MEAS}=\sqrt{2} \sigma_\mathrm{REF}$, so a single measurement gives the resolution of both LGADs. For the reference, an unirradiated HPK2 W43 LGAD biased at -160 V, which has a timing resolution of $33.4 \pm 0.6$ ps, was used.

\begin{figure}[htbp]
\centering
\includegraphics[width=.45\textwidth]{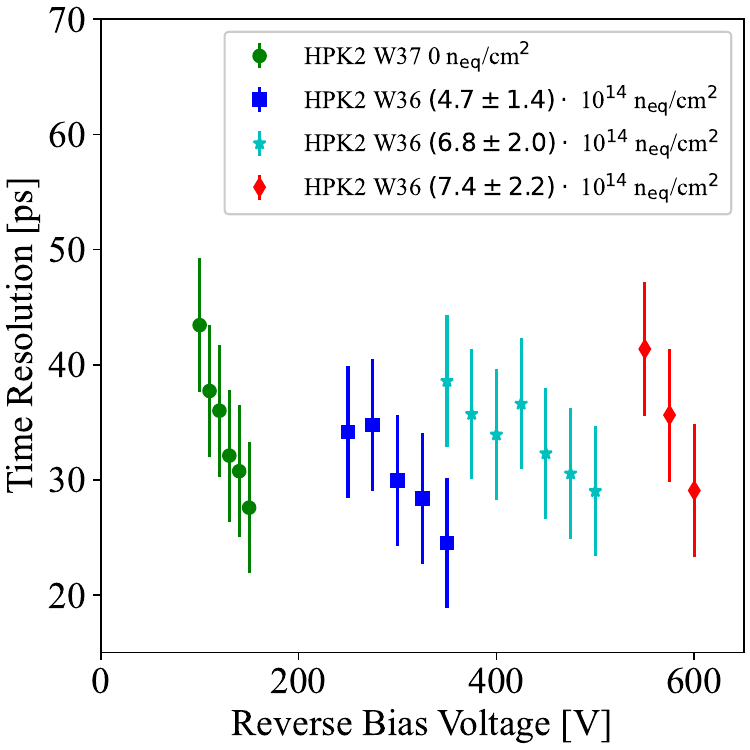}
\qquad
\includegraphics[width=.45\textwidth]{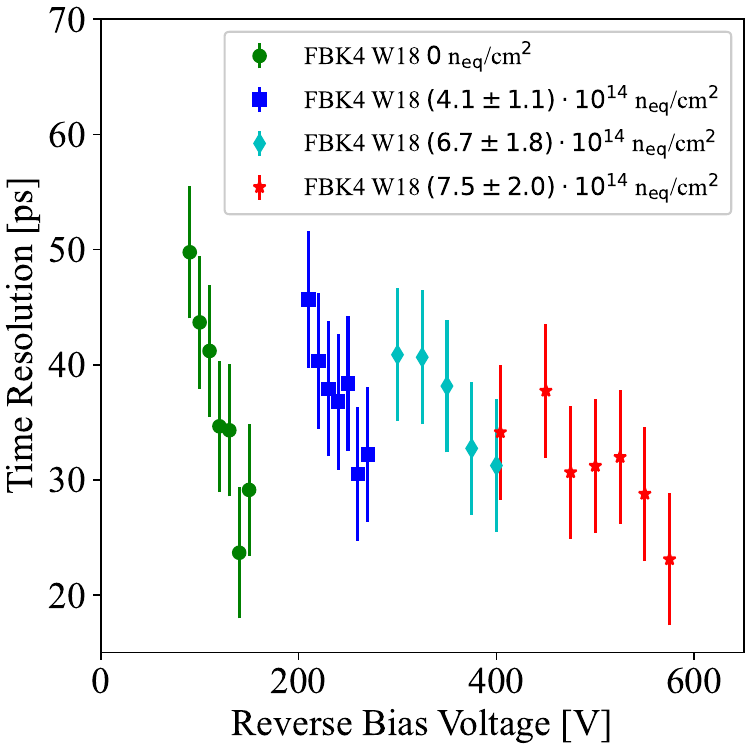}
\caption{Observed timing resolution of LGAD sensors processed as HPK2 (left) and FBK4 (right) up to the maximum applied fluence of $(7.4\pm2.2)\cdot10^{14} \mathrm{n}_\mathrm{eq} \mathrm{cm}^{-2}$ in 400 MeV and $(7.5\pm2.0)\cdot10^{14} \mathrm{n}_\mathrm{eq} \mathrm{cm}^{-2}$ 500 MeV in protons respectively.\label{fig:timing}}
\end{figure}

Figure \ref{fig:timing} shows an example measurement of the timing precision for FBK4 W18 and HPK2 W36 and W37. Both sensors' timing resolution improves with bias voltage, until near breakdown. The decrease in charge collection with fluence requires application of higher bias voltages to achieve comparable timing resolution from the irradiated sensors. For HPK2 W36 irradiated up to ~$(7.4 \pm ~2.2)\cdot10^{14} \mathrm{n}_\mathrm{eq} \mathrm{cm}^{-2}$ and for FBK4 W18 irradiated up to $(7.5 \pm 2.0)\cdot10^{14} \mathrm{n}_\mathrm{eq} \mathrm{cm}^{-2}$, the timing resolution does not exceed 50 ps. FBK4 W18 sensors were also irradiated to   $(11.0 ~\pm~ 2.4)~\cdot~10^{14} \mathrm{n}_\mathrm{eq} \mathrm{cm}^{-2}$ and $(15.7 \pm 3.3)\cdot10^{14} \mathrm{n}_\mathrm{eq} \mathrm{cm}^{-2}$ but no charge collection was measured in these sensors below 600~V. 

The systematic uncertainties on the timing resolution measurements were evaluated as follows. For all measurements, the oscilloscope was set to 100 mV/div (4 mV/bin). By comparing otherwise identical measurements with 200 mV/div and 70 mV/div, it was determined that the
choice of vertical digitization leads to a systematic error of 5 ps. The horizontal digitization was investigated by measuring at 5, 10, and 20 GSamples/s; from this study it was determined that the choice of horizontal bin width leads to a 2 ps uncertainty. The slope from a linear fit on timing resolution measurements at temperatures 7$^\circ$C  above and 7$^\circ$C  below the nominal temperature of -$30^\circ$C  was multiplied by the environmental chamber’s $\pm 1.5^\circ$C uncertainty, to infer a 2 ps uncertainty associated with temperature instability. The sources of systematic uncertainty are added in quadrature with the statistical error, and that total error is shown in Figure \ref{fig:timing}.

The charge collection was determined by recording the peak voltage from 2000 waveforms for each sensor for several bias voltages. The peak voltage distribution was fit to a Landau distribution convolved with a gaussian to represent noise. The most probable value (MPV) of the Landau distribution is related to the total charge collected. A 10 mV threshold (approximately 0.5 fC) was chosen for the measurements. If the lower tail of the signal’s Landau power spectrum fell below this threshold or if the electronic noise gaussian’s tail exceeded the most probable value of the signal’s Landau power spectrum, the data were discarded. 

\begin{figure}[htbp]
\centering
\includegraphics[width=.45\textwidth]{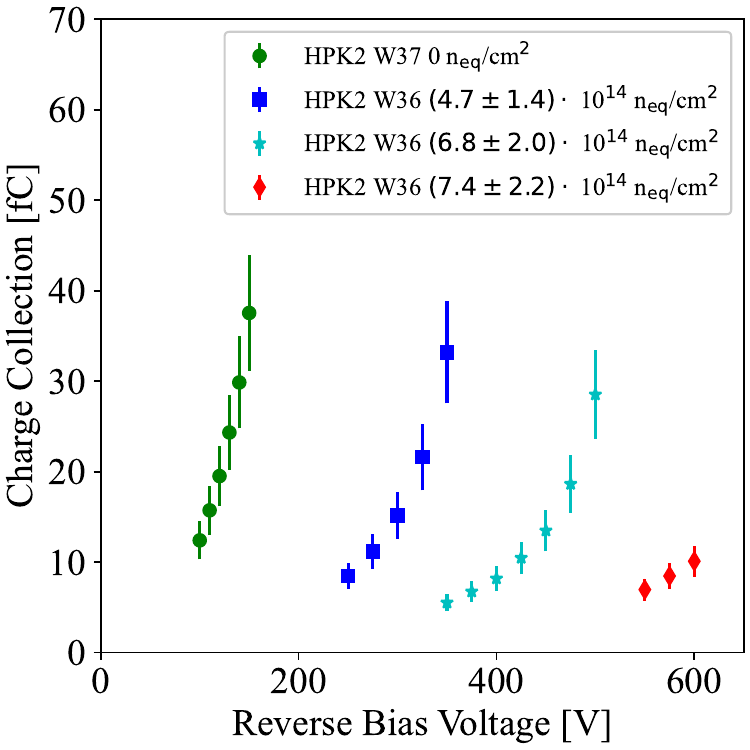}
\qquad
\includegraphics[width=.45\textwidth]{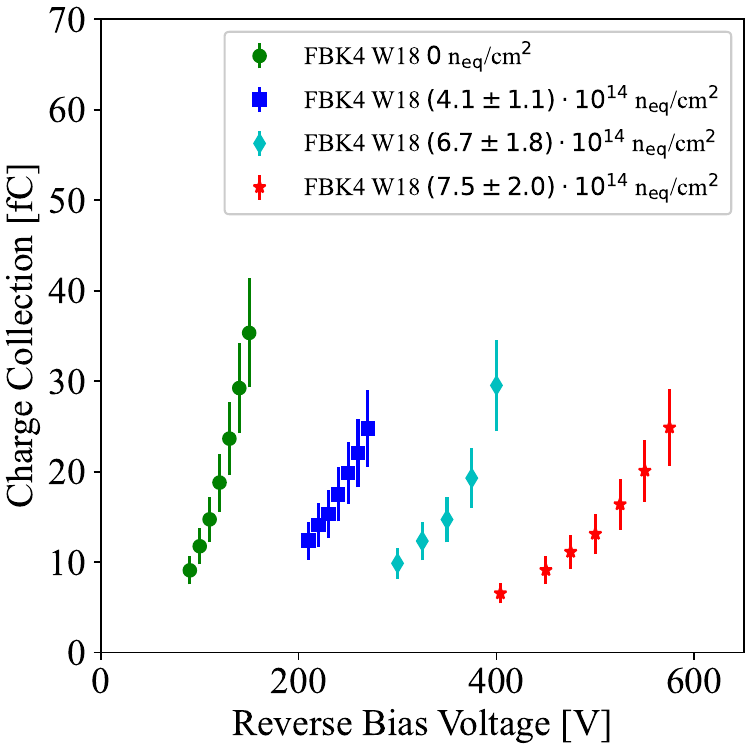}
\caption{Observed charge collection of LGAD sensors processed as HPK2 (left) and FBK4 (right) up to the maximum applied fluence of $(7.4\pm2.2)\cdot10^{14} \mathrm{n}_\mathrm{eq} \mathrm{cm}^{-2}$ in 400 MeV and $(7.5\pm2.0)\cdot10^{14} \mathrm{n}_\mathrm{eq} \mathrm{cm}^{-2}$ in 500 MeV in protons respectively.\label{fig:chargeCollection}}
\end{figure}

Figure \ref{fig:chargeCollection} shows the charge collection for FBK4 W18 and HPK2 W36 and W37. Charge collection above 30 fC is observed for HPK2 W36 up to $(6.8 \text{ }\pm \text{ }2.0)\cdot10^{14} \mathrm{n}_\mathrm{eq} \mathrm{cm}^{-2}$. For the sensor irradiated to $(7.4 \text{ }\pm  \text{ }2.2)\cdot10^{14} \mathrm{n}_\mathrm{eq} \mathrm{cm}^{-2}$, the charge collection is below 10 fC. For FBK4 W18, charge collection above 20 fC is observed up to $(7.5  \text{ }\pm \text{ } 2.0)\cdot10^{14} \mathrm{n}_\mathrm{eq} \mathrm{cm}^{-2}$. The FBK4 sensors irradiated to $(11.0 \pm 2.4)\cdot10^{14} \mathrm{n}_\mathrm{eq} \mathrm{cm}^{-2}$ and $(15.7 \pm 3.3)\cdot10^{14} \mathrm{n}_\mathrm{eq} \mathrm{cm}^{-2}$ had no measurable charge collection at bias up to 600 V and consequently they are not reported in Figures \ref{fig:timing} and \ref{fig:chargeCollection}. The charge collection was calibrated using a $0.30 \text{ } \pm \text{ } 0.05$ pF capacitor built into the dedicated preamp boards. It was found that for thin triangularly shaped pulses (like the LGAD signal), 1 fC corresponds to a 25 mV peak. Thus, measurements of the oscilloscope signal peak can be calibrated to universal units of charge collection. The error on the calibration capacitors’ capacitance, the error on the preamplifier board gain, and the statistical error on the fit are combined in quadrature and shown in Figure \ref{fig:chargeCollection}.

\subsection{Inter-electrode Resistance}

\begin{figure}[htbp]
	\centering
	\includegraphics[width=1\textwidth]{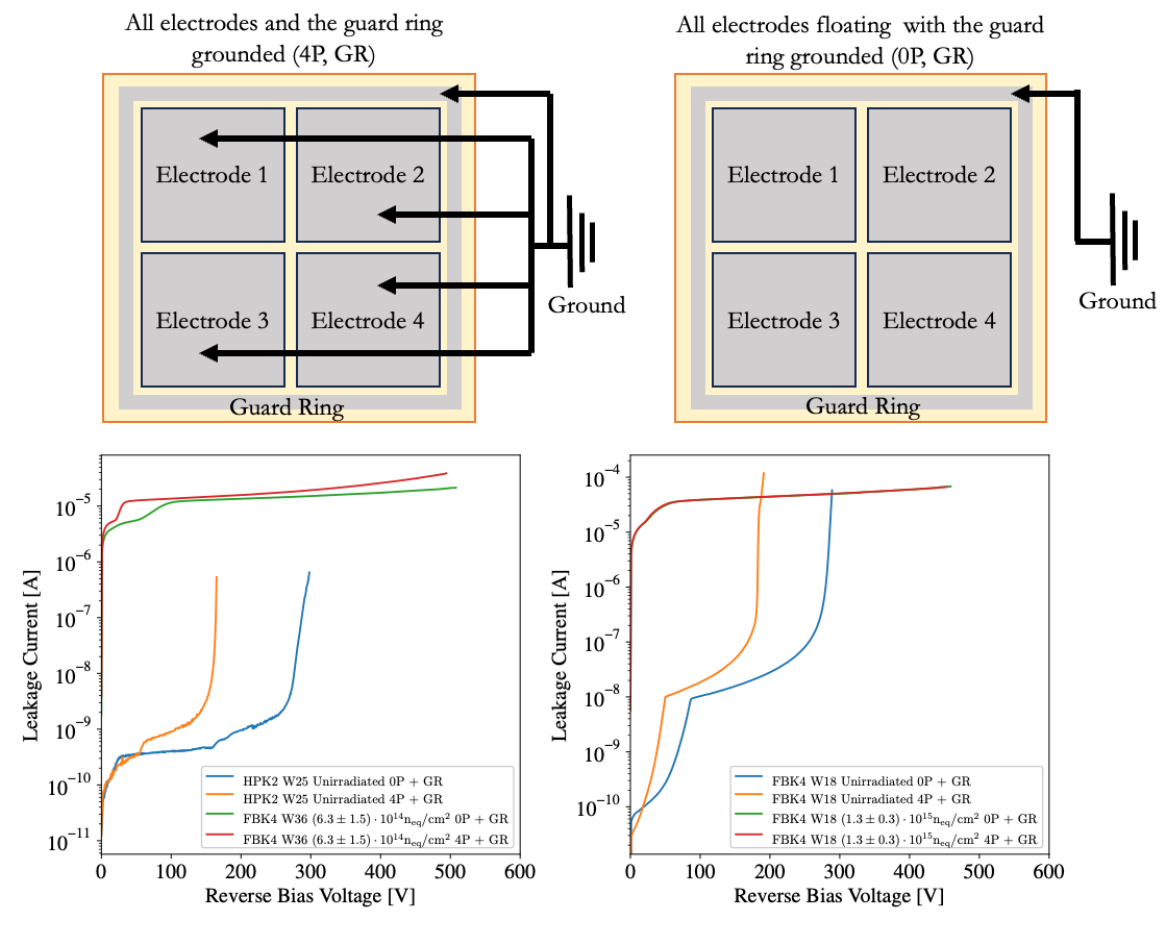}
	\caption{Schematic illustration of the leakage current measurement in the two modes: all four pads and the guard ring grounded (4P+GR, top left) and all four pads floating with the guard ring grounded (0P+GR, top right). Leakage current versus bias voltage is shown for HPK2 W25 and W36 (bottom left) and FBK4 W18 (bottom right) quads in these two modes, both un-irradiated and at the maximum fluence received. The difference in $V_\mathrm{gl}$, which corresponds to the first elbow in the curve, in these two modes is the punch-through voltage and is plotted in Figure \ref{fig:punchThrough}.  These plots are representative of all measured sensors.  \label{fig:quadIV}}
\end{figure}

If an electrode is floating, its potential is distributed to neighbors by punch-through. This process places a lower limit on the inter-electrode separation, for which the designer must anticipate the consequences in case a lost bump bond leads to breakdown at an electrode, which could then cascade to breakdown in neighbors. A study of leakage current versus bias voltage (IV), for the range of applied fluences, was carried out on the quad sensors to investigate the effects of proton irradiation on the inter-electrode isolation. Bias is applied to the back side of the sensor, and leakage current is measured with ground connected to the guard ring plus 0, 1, 2, 3, or all 4 pads. A sketch showing the grounding configuration and examples of this measurement for HPK2 W25 and W36 and FBK4 W18 are shown in Figure \ref{fig:quadIV}. The difference in $V_\mathrm{gl}$ between the 0 pads and 4 pads measurements is defined as the punch-through voltage, $\Delta V_\mathrm{gl}$. 

\begin{figure}[htbp]
\centering
\includegraphics[width=0.75\textwidth]{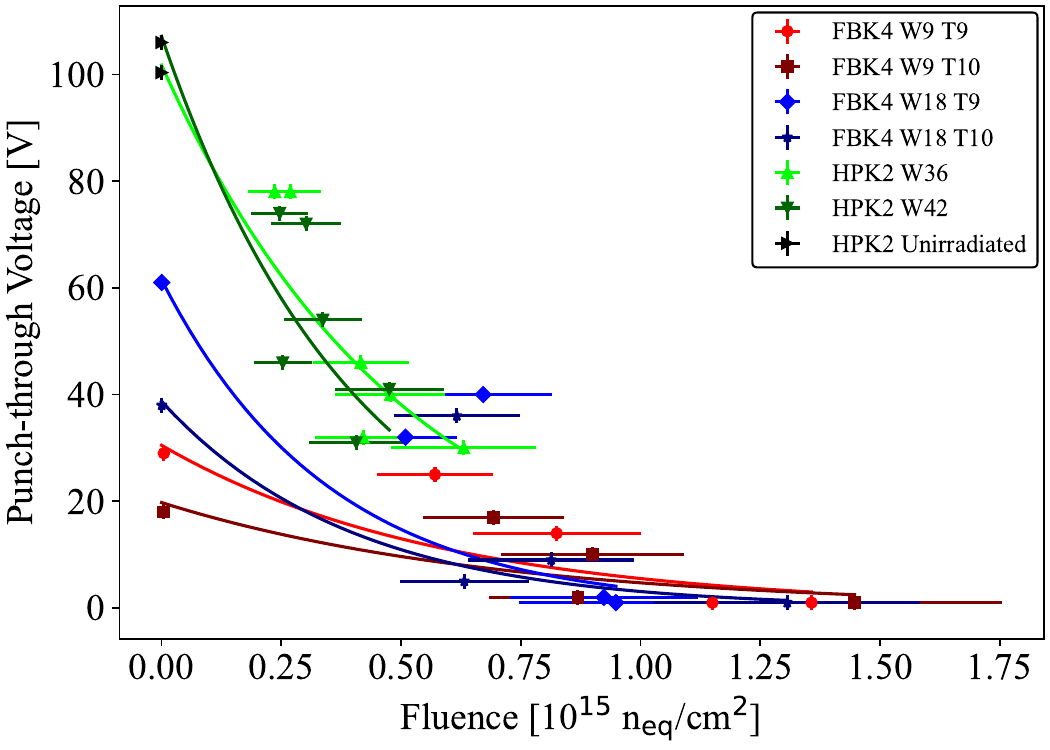}
\caption{Observed punch-through voltage, as a function of proton fluence applied to the FBK4 and HPK2 sensors. The inter-electrode spacing of all sensors is 50 $\mu$m. Each measurement is a different sensor. All measurements were performed at $20^\circ \mathrm{C}$. A decaying exponential function, ~$\Delta V_\mathrm{gl} (\phi)=~\Delta V_\mathrm{gl,0}\cdot ~e^{-k_\mathrm{punchThrough}\phi}$, is fit to the data represented by a solid line. The results of the fit are shown in Table \ref{tab:punchThrough}. \label{fig:punchThrough}}
\end{figure}

Figure \ref{fig:punchThrough} shows the punch-through voltage as a function of fluence. The horizontal error bars are the dose uncertainty and the vertical error bars are the quadrature sum of the error in $V_\mathrm{fd}$ with all 4 pads biased and with 0 pads biased. Punch-through between the guard ring and the pads occurs around 100 V for HPK2, and between 20 and 60 V for FBK4, prior to irradiation. The punch-through voltage of the FBK4 sensors decreases to nearly 0 after $10^{15} \mathrm{n}_\mathrm{eq} \mathrm{cm}^{-2}$. The HPK2 prototypes’ punch-through voltage also approaches zero at $5\cdot10^{14} \mathrm{n}_\mathrm{eq} \mathrm{cm}^{-2}$. In order to characterize the rate at which the punch-through approaches zero, a decaying exponential function, $\Delta V_\mathrm{gl} (\phi)=\Delta V_\mathrm{gl,0}\cdot e^{-k_\mathrm{punchThrough}\phi}$,  is fit to each measurement. The values of the exponential fit coefficients are in Table \ref{tab:punchThrough}. A larger punch-through decay constant, $k_\mathrm{punchThrough}$, indicates that the punch-through voltage decreases more readily with proton fluence. The drop in punch-through voltage indicates the loss of resistivity in the region between the pads and the guard ring. In addition, the $k_\mathrm{punchThrough}$ value is higher for the T9 quads than it is for the T10 quads from FBK. 


\begin{table}[htbp]
\centering
\caption{Extracted value of punch-through voltage decay constant, $k_\mathrm{punchThrough}$, from the fit function $\Delta V_\mathrm{gl} (\phi)=\Delta V_\mathrm{gl,0}\cdot e^{-k_\mathrm{punchThrough}\phi}$. The fits are shown in Figure \ref{fig:punchThrough}. \label{tab:punchThrough}}
\smallskip
\begin{tabular}{c|c} 
\hline
Wafer & $k_\mathrm{punchThrough}$ [10$^{-15}$$\mathrm{n}_\mathrm{eq}^{-1}$ cm$^{2}$] \\
\hline
FBK W9 T9 & $1.7\pm0.8$ \\
FBK W9 T10&   $1.4\pm0.8$  \\
FBK W18 T9&  $2.8\pm1.2$   \\
FBK W18 T10 & $2.5\pm 1.1$ \\
HPK W36&  $1.9\pm0.3$ \\
HPK W31 &  $2.5\pm0.4$ \\
\hline
\end{tabular}
\end{table}

\begin{figure}[htbp]
\centering
\includegraphics[width=0.75\textwidth]{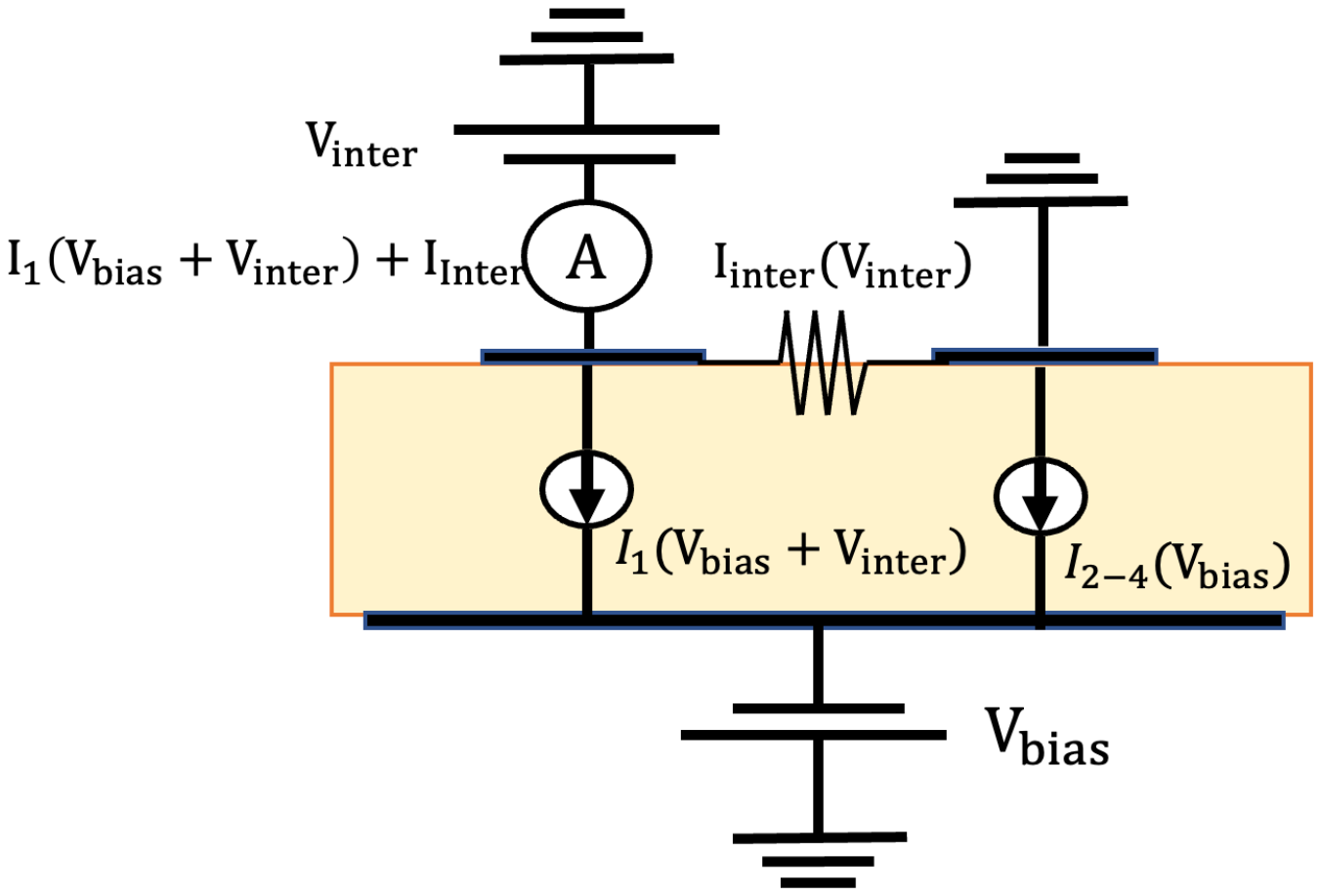}
\caption{Schematic of the inter-electrode resistance measurement setup. \label{fig:interPadResistance_schematic}}
\end{figure}

To further evaluate this effect, a direct measurement of the inter-electrode resistance was performed with the sensors in fully depleted mode. A schematic of how this measurement is set up is shown in Figure \ref{fig:interPadResistance_schematic}. The first electrode was connected to a Keithley 2410 source meter with a common ground. The bias, $\mathrm{V}_\mathrm{inter}$, on the Keithley 2410 allowed the first electrode to be biased differently from the remaining three electrodes and produced a reading of the current resulting from this bias. The remaining three electrodes and the guard ring were grounded. The bias, $\mathrm{V}_\mathrm{bias}$, on the backside of the sensor was set to -100 V so that the sensor was in fully depleted mode. $\mathrm{V}_\mathrm{inter}$ was varied from -2 V to 2 V in 0.5 V steps relative to ground. At each value of $\mathrm{V}_\mathrm{inter}$, the leakage current between the first electrode and the remaining electrodes and guard ring was measured. The net bias on the first electrode is $\mathrm{V}_\mathrm{bias} + \mathrm{V}_\mathrm{inter}$, which slightly increases the sensor's bulk leakage current for the higher magnitude bias measurements. This effect on the leakage current is not due to the inter-electrode characteristics, so it is subtracted by making complementary measurements with $\mathrm{V}_\mathrm{inter}$ set to 0 V and varying $\mathrm{V}_\mathrm{bias}$ from -102 V to -98 V.

\begin{figure}[htbp]
\centering
\includegraphics[width=0.75\textwidth]{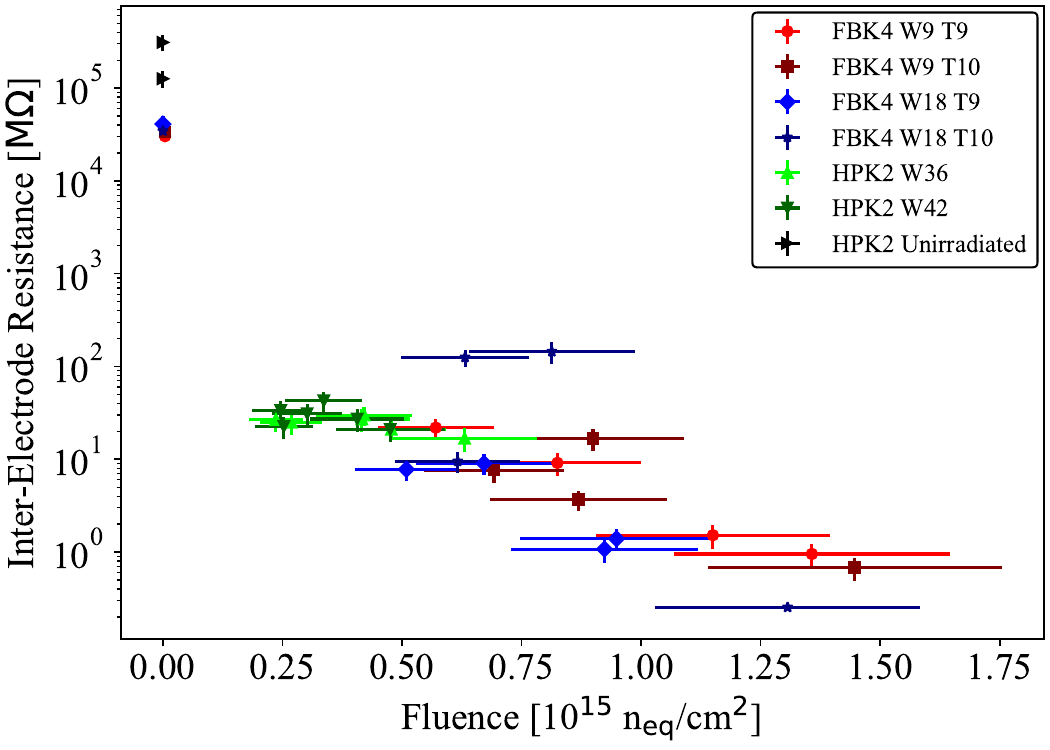}
\caption{Measured inter-electrode resistance between one biased electrode and the three adjacent grounded electrodes in a quad sensor for both manufacturers as a function of proton fluence. Each measurement is a different sensor. All measurements were performed at $20^\circ \mathrm{C}$. \label{fig:interPadResistance}}
\end{figure}

A linear fit to the corrected leakage current versus bias voltage was used to extract the resistance. The inter-electrode resistance, which is the resistance between the first electrode and all adjacent grounded features, versus applied fluence is plotted in Figure \ref{fig:interPadResistance}. The HPK2 LGADs have initial inter-electrode resistance of 10$^5$  M$\Omega$, and it drops to $\sim$10 M$\Omega$  after $5\cdot10^{15} \mathrm{n}_\mathrm{eq} \mathrm{cm}^{-2}$. The FBK4 LGADs have initial inter-electrode resistance of $5\cdot 10^4$  $\mathrm{M}\Omega$, and it drops to $\sim$1 $\mathrm{M}\Omega$ after $10^{15} \mathrm{n}_\mathrm{eq} \mathrm{cm}^{-2}$. 

A charge collection measurement was performed on the quad LGADs with a 1060 nm pulsed laser. One of the sensors on the quad was wire bonded to the dedicated readout board. The other three sensors were left floating. The laser was focused into the optical window of the wire bonded sensor and the laser pulse was tuned until the collected signal was comparable in shape and amplitude to a waveform from the $^{90}$Sr source. Then the floating LGADs were stimulated with the laser. No signal was observed on the bonded sensor due to the stimulated adjacent floating pads even at biases exceeding the punch-through voltage. This was the case for all fluences and indicates that the pads are still well isolated for fast signals after the maximum proton dose has been received. 

\section{Conclusions and Outlook}

Effects due to radiation damage from 400 and 500 MeV protons applied to HPK2 and FBK4 LGADs respectively have been characterized through leakage current, capacitance, charge collection, and timing resolution measurements. The damage caused by these protons to the inter-electrode regions has been characterized with measurements of the punch-through voltage, direct measurement of inter-electrode resistance with DC, and charge collection measurements from adjacent, floating sensors. 

The proton damage to HPK2 LGADs, which lack carbon co-implantation in the gain layer, appears greater in terms of the acceptor removal constant. The measured acceptor removal constant for the HPK2 LGADs is between $\sim 6\cdot10^{-16} \mathrm{cm}^2/\mathrm{n}_\mathrm{eq}$ and $\sim 8\cdot10^{-16} \mathrm{cm}^2/\mathrm{n}_\mathrm{eq}$ whereas the acceptor removal constant for the FBK4 LGADs is between $\sim 3\cdot10^{-16} \mathrm{cm}^2/\mathrm{n}_\mathrm{eq}$ and $\sim 6\cdot10^{-16} \mathrm{cm}^2/\mathrm{n}_\mathrm{eq}$. The superior response of the FBK4 LGADs may be due to the carbon co-implantation in the gain layer~\cite{effectOfCarbon}. 

Compared with neutron damage reported in \cite{acceptorRemovalNeutrons}, the acceptor removal constant from proton damage is between 2 and 3 times greater for the FBK4 LGADs and around 1.5 times greater for the HPK2 LGADs. This is suggestive of differences in the damage mechanisms associated with charged and neutral radiation that are not fully accounted for in the NIEL scaling hypothesis \cite{NIELS} used to normalize the damage.

The extracted damage constant values reported in Table \ref{tab:damageConstant} indicate that the leakage current increased more rapidly in the FBK4 wafers than in the HPK2 wafers. The damage constant extracted from the pin sensors is consistent with the LGAD damage constant for both HPK2 and FBK4, so the difference in damage constant in not likely due to the gain layer properties.

Sensors from both wafers show exceptional timing resolution and charge collection for moderate proton fluences. The HPK2 W36 and FBK4 W18 sensors maintain a timing resolution better than 35 ps after $(7.4\text{ }\pm\text{ }2.2)\cdot10^{14} \mathrm{n}_\mathrm{eq} \mathrm{cm}^{-2}$ and $(7.5\text{ }\pm\text{ }2.0)\cdot10^{14} \mathrm{n}_\mathrm{eq} \mathrm{cm}^{-2}$ respectively. The charge collection of the HPK2 W36 was initially about 30 fC and dropped to around 10 fC after $(7.4\text{ }\pm\text{ }2.2)\cdot10^{14} \mathrm{n}_\mathrm{eq} \mathrm{cm}^{-2}$. The charge collection of the FBK4 W18 remains above 20 fC after $(7.5\pm2.0)\cdot10^{14} \mathrm{n}_\mathrm{eq} \mathrm{cm}^{-2}$. The FBK W18 sensors irradiated to $(11.0 \pm 2.4)\cdot10^{14} \mathrm{n}_\mathrm{eq} \mathrm{cm}^{-2}$ and $(15.7 \pm 3.3)\cdot10^{14} \mathrm{n}_\mathrm{eq} \mathrm{cm}^{-2}$  had no measurable charge collection. 

The inter-electrode characteristics of the FBK4 and HPK2 sensors are comparable. After irradiation, both devices’ punch-through voltages approach zero. The FBK4 inter-electrode resistance drops to $\sim1$ $\mathrm{M} \Omega$ after $10^{15} \mathrm{n}_\mathrm{eq} \mathrm{cm}^{-2}$ while the HPK2 inter-electrode resistance drops to $\sim10$ $\mathrm{M}\Omega$  after $5\cdot10^{14} \mathrm{n}_\mathrm{eq} \mathrm{cm}^{-2}$. Both remain well isolated for fast signals.





\acknowledgments

The opportunity to use the LANSCE facility, made possible by Steven Wender and Kranti Gunthoti of Los Alamos National Laboratory, is gratefully acknowledged. The Fermilab Irradiation Test Area facility, led by Evan Niner, was also critical to this work.  Support from Paulo Oemig of New Mexico State University and the encouragement of Jeremy Perkins and Regina Caputo, both of NASA/GSFC, are deeply appreciated. Carl Willis of the UNM Department of Nuclear Engineering provided essential help with the dosimetry.  This work was made possible by support from U.S. Department of Energy grant DE-SC0020255 and from ARIS grant J7-4419, Slovenia. It was also supported by the National Aeronautics and Space Administration (NASA) under Federal Award Number 80NSSC20M0034 (2020-RIG).




\begin{thebibliography}{99}

\bibitem{HGTD}
M.P. Cassado,
\emph{A High Granularity Timing Detector for the ATLAS Phase-II Upgrade},
Nucl. Inst. and Meth. A 1032 (2022) 166628.


\bibitem{ETL}
CMS Collaboration,
\emph{A MIP Timing Detector for the CMS Phase-2 Upgrade}, 
Tech. Rep., CERN, Geneva, 2019, URL https://cds.cern.ch/record/2667167.

\bibitem{LGAD1}
G. Pellegrini et al., 
\emph{Technology developments and first measurements of Low Gain Avalanche Detectors (LGAD) for high energy physics applications}, 
Nucl. Instrum. Meth. A765 (2014) 12 -16.

\bibitem{LGAD2}
M. Carulla et al., 
\emph{First 50 $\mu \mathrm{m}$ thick LGAD fabrication at CNM}, 
28th RD50 Workshop Torino, Italy, June 7th 2016.

\bibitem{breakdown}
N. Moffat et al. 
\emph{Low Gain Avalanche Detectors (LGAD) for particle physics and synchrotron applications},
2018 JINST 13 C03014.


\bibitem{gainDiff}
H.F.-W. Sadrozinski et al.,
 \emph{4D tracking with ultra-fast silicon detectors}, 
 2018 Rep. Prog. Phys. 81, 026101.
 
 \bibitem{carbonEffect}
M. Ferrero et al. 
\emph{Radiation resistant LGAD design},
Nucl. Inst. and Meth. A 919 (2019), https://doi.org/10.1016/j.nima.2018.11.121.

\bibitem{UFSD_c_table}
M. Ferrero, et al., \emph{An Introduction To Ultra-Fast Silicon Detectors}, CRC Press, 2021, p. 128, http://dx.doi.org/10.1201/9781003131946.

\bibitem{diffusion}
M. Ferrero, et al.,
 \emph{A summary of the radiation resistance of carbonated gain implants},
37th RD50 Workshop, Zagreb, Croatia, November 19th, 2020, https://indico.cern.ch/event/896954/contributions/4106307.

\bibitem{MCNP}
C.J. Werner et al.,
\emph{MCNP6.2 Release Notes},
Los Alamos National Laboratory, 
report LA-UR-18-20808, (2018).

\bibitem{Morgan}
G.L. Morgan et al.,
\emph{Total cross sections for the production of ~$^{22}$Na and $^{24}$Na in proton-induced reactions on $^{27}$Al from 0.40 to 22.4 GeV},
Nucl. Instr. Meth. B. 211 (2003) 297-304.

\bibitem{Hoeferkamp}
M.R. Hoeferkamp, A. Grummer, I. Rajen, and S. Seidel,
\emph{Application of p-i-n photodiodes to charged particle fluence measurements beyond $10^{15}$ 1-MeV-neutron-equivalent/cm$^2$},
Nucl. Instr. and Meth. A 890 (2018) 108-111.

\bibitem{singleEventBurnout}
G. Laštovička-Medin et al. 
\emph{A brief overview of the studies on the irreversible breakdown of LGAD testing samples irradiated at the critical LHC-HL fluences},
JINST 17 C07020 (2022).

\bibitem{damageConstant}
G. Lindström, M. Moll and E. Fretwurst,
\emph{Radiation hardness of silicon detectors: A challenge from high-energy physics}, 
Nucl. Instrum. Meth. A 426 (1999) 1.

\bibitem{effectOfCarbon}
M. Moll
\emph{Radiation Damage in Silicon Detectors}
CERN EP-TA1-SD Seminar 14.2.2001

\bibitem{vglEquation}
M. Ferrero, et al., \emph{An Introduction To Ultra-Fast Silicon Detectors}, CRC Press, 2021, p. 97, http://dx.doi.org/10.1201/9781003131946.

\bibitem{damageMechanism}
M. Moll. 
\emph{Acceptor removal – Displacement damage effects involving the shallow acceptor doping of p-type silicon devices},
PoS (Vertex2019) 027.


\bibitem{radEffectLGAD}
G. Kramberger et al. 
\emph{Radiation effects in Low Gain Avalanche Detectors after hadron irradiations}, 
2015 JINST 10 P07006.


\bibitem{neutronLGAD}
Y. Jin et al. 
\emph{Experimental Study of Acceptor Removal in UFSD}, 
Nucl. Inst. and Meth. A 983 (2020) https://doi.org/10.1016/j.nima.2020.164611.


\bibitem{UCSCboard}
N. Cartiglia et al.
\emph{Beam test results of a 16 ps timing system based on ultra-fast silicon detectors}, 
Nucl. Inst. and Meth. A 850 (2017) 83, arXiv:1608.08681.


\bibitem{acceptorRemovalNeutrons}
A. Howard et al. 
\emph{LGAD measurements from different producers}
37th RD50 Workshop (2020). 

\bibitem{NIELS}
A. Vasilescu and G. Lindström 
\emph{Notes on the fluence normalisation based on the NIEL scaling hypothesis}, 
ROSE/TN/2000-02, 


\end{thebibliography}


\end{document}